\documentclass[preprint2]{aastex}

\shorttitle{Globular cluster integrated spectra}
\shortauthors{Coelho et al.}

\begin{document}


\title{Chemical abundance anticorrelations in globular cluster stars: The effect on cluster integrated spectra}


\author{P. \ Coelho\altaffilmark{1}}
\affil{N\'ucleo de Astrof\'{\i}sica Te\'orica, Universidade Cruzeiro do Sul, 
R. Galv\~ao Bueno 868, Liberdade, 01506-000, S\~ao Paulo, Brasil; paula.coelho@cruzeirodosul.edu.br}

\author{S. M.\ Percival\altaffilmark{2}, M. \ Salaris\altaffilmark{2}}
\affil{Astrophysics Research Institute, Liverpool John Moores
University, 12 Quays House, Birkenhead, CH41 1LD, UK; smp,ms@astro.livjm.ac.uk}



\begin{abstract}
It is widely accepted that individual Galactic globular clusters harbor two coeval 
generations of stars, the first one born with the `standard' $\alpha$-enhanced metal mixture 
observed in field Halo objects, the second one characterized by an anticorrelated CN-ONa abundance 
pattern overimposed on the first generation, $\alpha$-enhanced metal mixture. 
We have investigated with appropriate stellar population synthesis models
how this second generation of stars affects the integrated spectrum of a typical metal rich 
Galactic globular cluster, like 47\,Tuc,
focusing our analysis on the widely used Lick-type indices.
We find that the only indices appreciably affected by the abundance anticorrelations are 
Ca4227, G4300, ${\rm CN_1}$, ${\rm CN_2}$ and NaD. The age-sensitive Balmer line,  
Fe line and the [MgFe] indices widely used to determine age, 
Fe and total metallicity of extragalactic systems are largely insensitive to the 
second generation population.
Enhanced He in second generation stars 
affects also the Balmer line indices of the integrated spectra,  
through the change of the turn off temperature and -- in the assumption that the mass loss 
history of both stellar generations is the same  -- the 
horizontal branch morphology of the underlying isochrones.
\end{abstract}



\section{Introduction}

A large body of spectroscopic data published in the last 10 years has conclusively established 
the existence of primordial surface chemical abundance variations of C, N, O, Na  
-- and sometimes of Mg and Al --  in individual Galactic 
globular clusters (GCs) of all metallicities, from 
very metal rich objects like NGC6441, to the most metal poor ones like M15 \citep[see, e.g.,][]{cbg10}. 
When considering luminosities below approximately the red giant branch (RGB) bump level,
within a given GC the chemical abundance patterns appear to be roughly 
the same all along the RGB down to the turn-off (TO) region 
\citep[see, i.e.][]{g01, cm05},  
and show CN and ONa (and sometimes also MgAl) anticorrelations 
for about 2/3 of the stars, 
superimposed onto a standard (i.e. consistent with the distribution found in field halo stars) 
$\alpha$-enhanced heavy-element distribution ([$\alpha$/Fe]$\sim$ 0.3--0.4). 
The anticorrelations display a range of negative variations of C and O accompanied by increased N and Na 
abundances, whose extension varies from cluster to cluster \citep{c05, c09}. 
The remaining 1/3 of cluster stars 
shows an homogeneous standard $\alpha$-enhanced metal mixture.
At luminosities brighter than the RGB bump, observations point to the efficiency of some deep mixing 
process that decreases the surface C and increases N in all cluster stars \citep{yong, gcs04, sneden},   
similarly to the case of field Halo stars \citep{g00}, whereas the ONa pattern is essentially unchanged 
compared to lower luminosities. This effect varies from cluster to cluster,  
with a cluster like M13 depleting C much more efficiently than others \citep{sm, sneden}.

The accepted working scenario to explain these CNONa anticorrelations 
prescribes that the stars currently evolving in a GC were born with the observed CNONa patterns. 
Intermediate-mass asymptotic giant branch (AGB) stars in the range $\sim 3-8 M_{\odot}$ 
(and/or the slightly more massive 
super-AGB stars) and massive rotating stars are considered viable sources of the necessary heavy-element 
pollution \citep[see, e.g.,][]{vd05, decr07}. 
Slow winds from the envelopes of AGB (or super-AGB stars) or from the equatorial disk formed around fast rotating 
massive stars inject matter into the intra-GC medium after a time of order $10^6 - 10^8$~yr, depending on 
the polluter. Provided that a significant fraction of the material is not lost 
from the cluster, new stars (second generation, but essentially coeval with the 
polluters' progenitors, given their short evolutionary timescales) may be able to form directly out of pristine gas  
polluted to varying degrees by these ejecta, that will show the observed anticorrelation patterns.
Another  byproduct of this pollution may possibly be 
an enhanced initial He-abundance for these second generation stars. 
Recent numerical models by \citet{derc08, derc10} have started to explore in quantitative details this broad picture. 

If Galactic GCs are the local counterpart of extragalactic globulars, one should  
expect similar abundance anticorrelations to be imprinted in the   
integrated spectra of GCs populating the haloes of external galaxies. 
Given that GCs are employed as tracers of the star formation histories of spheroids 
\citep[see, e.g. the review by][]{bs06}, and their spectra are usually modelled considering a   
simple stellar population (SSP) with a single initial chemical composition (and age), 
it is important to assess to what extent these abundance anticorrelations modify our intepretation of GC 
integrated spectra when compared to the case of a SSP. 
Also, integrated spectra of Galactic GCs are often used to test SSP models 
\citep[see, i.e.,][]{lw05, basti09, walcher+09, vazd10, tjm} 
and the interpretation of these comparisons is clearly dependent on the effect 
of these abundance anticorrelations on the cluster spectra.

We address for the first time these questions, focusing mainly 
on the predicted Lick-type absorption 
feature indices, that are routinely employed to determine ages and chemical compositions of extragalactic 
stellar populations 
\citep[see, e.g.,][and references therein]{burstein, w94, trager, pfb04, puzia, thomas05, gs08}.
The next section presents our methodology and theoretical modelling, and is followed by  
an analysis of the results and a discussion. 
  
\section{Models}\label{models}

We have started our analysis by considering two $\alpha$-enhanced SSPs  -- metal mixture with 
[$\alpha$/Fe]$\sim$ 0.4 -- with ages t=12 and 14 Gyr, 
[Fe/H]=$-$0.7 ($Z$=0.008) and initial He mass fraction $Y$=0.256, 
as representative of the first generation chemical composition in typical metal rich 
Galactic GCs, like 47~Tuc, whose integrated spectrum is often used for testing stellar population synthesis models.
For both ages we have then considered a second generation population whose metal composition has  
C decreased by 0.30~dex, N increased by 1.20~dex, O decreased by 0.45~dex and Na increased by 0.60~dex with 
respect to the first generation $\alpha$-enhanced mixture, all other metal abundances  
being unchanged. This pattern is typical of values close to the upper end of the observed anticorrelation 
patterns in Galactic GCs \citep{c05, c09}. 
The metal distribution of this second generation coeval population has the same C+N+O sum 
and the same Fe abundance (as a consequence also the total metallicity $Z$ will be practically the same) 
as the first generation composition, in agreement with spectroscopic measurements on 
second generation stars within individual Galactic GC\footnote{A well known exception to the constancy of Fe within a 
Galactic GC is $\omega$ Cen, that displays a large range of Fe abundances in addition to He enhancements and 
CNONa anticorrelations \citep[see, e.g.,][and reference therein]{b10}.}.

For both the 12 and 14~Gyr second generation populations we have accounted for two alternative 
values of $Y$, e.g. $Y$=0.256 -- as in the first stellar generation -- and $Y$=0.300, to include a 
possible enhancement of He in second generation stars. This latter 
choice is consistent with constraints on the typical enhancement of He in Galactic GCs, 
as determined by \citet{bragag10}. 
We stress that these choices 
for the chemical composition of second generation populations are very general, and 
should give us a realistic estimate of their impact on GC analyses using SSPs. 
The accurate modelling of individual clusters would require, however, much more specific prescriptions, 
given that  
the distribution of CNONa abundances in second generation stars varies on a cluster-to-cluster basis. 
One would need to implement the effect of a continuous variations of CNONa elements up to the maximum 
values of the anticorrelations -- that is cluster dependent -- plus the exact number distribution of stars along 
the anticorrelation pattern, also cluster dependent. Even this would not help a detailed comparison with 
extragalactic GCs, given that at present we do not know -- nor we can predict -- the
distribution of the abundance anomalies in these objects. 
The effect of the deep mixing along the upper RGB that modifies the CN abundances (that 
to some extent seems, again, cluster-dependent) 
is not accounted for in our calculations (nor in any other existing calculations of integrated spectra of GCs), 
but its effect goes in the same direction of the CN primordial anticorrelation, for it tends to amplify the 
range of C depletions and N enhancements. 

Overall, our selection of representative extreme values of the CNONa variations 
will provide a first important indication of the maximum effect
of these abundance anomalies on GC integrated spectra, and will serve as a guideline to interpret 
the abundance pattern derived from fitting SSPs to the observed spectra of Galactic and extragalactic GCs.
We have considered in this
first investigation on the subject one single [Fe/H] value, 
typical of the metal rich subpopulations of Milky Way -- and of 47Tuc, whose integrated spectrum 
is a benchmark for population synthesis models. 
The sensitivity of the spectra 
to the abundance anomalies may be quantitatively different when considering different [Fe/H] values.

For each of the first generation SSPs we have then calculated synthetic integrated spectra as follows. 
The underlying isochrones are the BaSTI $\alpha$-enhanced isochrones (with Reimers mass loss parameter 
$\eta$=0.2) by \citet{basti06}, with [Fe/H]=$-$0.7 ($Z$=0.008). 
The 12 and 14~Gyr isochrones representing the second generation populations are the same as the 
first generation 
$\alpha$-enhanced ones with the appropriate $Y$ abundance (taken from the BaSTI database), 
for we have verified with additional stellar model calculations that the effect of the two   
different metal mixtures on the isochrones is negligible, provided that the Fe abundance 
and the C+N+O sum are unchanged \citep[see also][]{swf06}.

We have then computed a grid of 57 
stellar synthetic spectra for both first and second generation metal mixtures,  
that cover the gravity-effective temperature parameter space spanned by the corresponding 12 and 14~Gyr isochrones. 
The model atmospheres have been calculated using the code 
ATLAS12 \citep{kur81,kur05,cas05}. Spectra were then computed with the code SYNTHE
\citep{kur81,sbo04}, for the wavelength region 3500 to 6000\,${\rm \AA}$ and 
convolved to a spectral resolution $R=\lambda/\delta\lambda=10000$. 
The atomic line list adopted in the computations is based on the compilations by
\cite{coe05} and \cite{cash04}. Lines for the molecules C2, CH, CN,
CO, H2, MgH, NH, OH, SiH and SiO from \cite{kur93} were included for all stars, 
and TiO lines from \cite{schwenke98} were included for stars cooler than 4500~K. 

From the appropriate isochrone and grid of synthetic stellar spectra  
we have calculated the integrated spectrum for these first and second generation populations, 
employing the \citet{kr01} initial mass function, and computed the corresponding values  
$I$ of the standard Lick-type indices as defined in Table~2 of \citet{trager}, plus the 
${\rm H\gamma}$ and ${\rm H\delta}$ indices defined in \citet{wo97}. 
Of the 21 indices in \citet{trager} we had to neglect Ti${\rm O_1}$ and 
Ti${\rm O_2}$, that are beyond the upper limit of the wavelength range of our spectra. 
The values of $I_{1st \ generation}$ and $I_{2nd \ generation}$  
have been determined as in \citet{basti09}, using the LECTOR program by A. Vazdekis\footnote{available
at \url{ http://www.iac.es/galeria/vazdekis/index.html}}
and measured directly on our high resolution spectra, i.e., they are not 
transformed onto the Lick system, for 
the community is rapidly moving to measure indices at higher resolution than the original Lick 
system, especially since the publication of models employing the MILES spectral library 
\citep{miles, vazd10}. 
As a test, we have degraded the resolution of our spectra to 2.3 \AA, the resolution of the MILES 
spectral library and repeated the analysis described below. Although the 
absolute values of the indices have a different zero point, the size of the differential effects 
due to varying metal mixtures (and He abundances) are unchanged.

%%%%%%%%%%%%%%%%%%%%%%%%%%%%%%%%%%%%%%%%%%%%%%%%%%%%%%%%%%%
%                            Tables
%%%%%%%%%%%%%%%%%%%%%%%%%%%%%%%%%%%%%%%%%%%%%%%%%%%%%%%%%%%

\begin{table*}
\begin{center}
\caption{Variations of the values of the Lick-type indices for our 14~Gyr old 
first and second generation populations, both with $Y$=0.256, and the corresponding 
$\Delta$[Fe/H] differences (see text for details) .\label{tbl1}}
\begin{tabular}{rrr}
\tableline\tableline
Index & $\Delta I$ & $\Delta$[Fe/H] \\
\tableline
${\rm H\delta_F}$ &               $-$0.068&    $--$ \\
${\rm H\gamma_F}$ &                  0.274&    $--$ \\
${\rm CN_1}$      &                  0.084&    1.960 \\
${\rm CN_2}$      &                  0.087&    1.324 \\
Ca4227   &                        $-$0.651& $-$0.306 \\
G4300    &                        $-$0.572& $-$0.360 \\
Fe4383   &                        $-$0.073& $-$0.022 \\
Ca4455   &                           0.000& $-$0.000 \\
Fe4531   &                        $-$0.014&  $-$0.006 \\
${\rm C_2}$4668 &                 $-$0.115& $-$0.063 \\
${\rm H\beta}$   &                $-$0.032&    $--$ \\
Fe5015   &                           0.056&    0.016 \\
${\rm Mg_1}$      &                  0.000&    0.000 \\
${\rm Mg_2}$     &                $-$0.004& $-$0.016 \\
Mgb     &                         $-$0.158& $-$0.039 \\
Fe5270  &                            0.004&    0.002 \\
Fe5335  &                            0.010&    0.005 \\
Fe5406  &                         $-$0.005&  $-$0.004 \\ 
Fe5709  &                         $-$0.060&  $-$0.068\\ 
Fe5782  &                            0.012&     0.022 \\
NaD     &                            1.346&     0.733 \\
\tableline
\end{tabular}
\tablecomments{Units are \AA \ but for the CN1, CN2 indices, whose values 
are expressed in magnitudes.}
\end{center}
\end{table*}

To assess the impact of second generation stars on ages and chemical composition of unresolved GCs, 
we have chosen to compare 
%$I_{2nd \ generation}$ 
our predictions with the values from the grid 
of $\alpha$-enhanced SSPs by \citet{basti09}, that spans a wide range of ages and metallicities. 
These spectra have been calculated employing the same BaSTI $\alpha$-enhanced isochrones 
(with Reimers mass loss parameter 
$\eta$=0.2) used in this paper, the same \citet{kr01} initial mass function, and the  
synthetic stellar spectral library by \citet{munari}.
These SSP spectra we adopted 
%have computed what we denote 
as the source of a 'reference' grid of values 
$I$ of the indices for first generation SSPs with [Fe/H] between $-$1.84 and +0.05, and age t between 1.25 and 14~Gyr.
We found small offsets in the reference index values for the 12 and 14~Gyr, [Fe/H]=$-$0.7, 
$\alpha$-enhanced SSPs, when compared to our new ATLAS12 calculations,  
due to differences in the stellar spectral libraries. To ensure these small offsets do not play any 
significant role in our analysis, we have decided to use our new results in a purely differential way. 
For each of the selected ages and 
$Y$ of second generation stars, we have determined from our calculations the differences 
$\Delta I$=$I_{2nd \ generation}-I_{1st \ generation}$. 
The differences $\Delta I$  have been then added to the 
corresponding index values for the 12 and 14~ Gyr 'reference' 
(first generation) $\alpha$-enhanced grid, to represent the corresponding coeval second generation.  
These are the index values employed in the analysis that follows.

\section{Results and discussion}\label{results}

In the following we compare the values of 
$I$ for the various indices in first and second generation populations with the same age, 
to quantify how Lick-type indices respond to the presence of a coeval 
second stellar generation with CNONa anticorrelations in individual GCs.
It is important to stress from the outset that all comparisons that follow give a quantitative estimate of the 
{\sl maximum} effect on the Lick-type indices of GCs, for we are comparing a pure first generation with a 
pure second generation population with extreme values of the CNONa anticorrelations. In a real GC 
the distribution of CNONa abundances ranges (most probably because of the effect of dilution with pristine gas) 
from the values of the standard first generation 
$\alpha$-enhanced mixture, to extreme values like the ones used in our modelling (and also, the specific 
properties of the CNONa distribution vary from cluster to cluster). One expects therefore that {\sl in real GCs  
the variations $\Delta I$ will attain values somewhere within the range derived from our analysis}.

We have first analyzed the case of the same initial He-abundance in first and second generation stars. 
We have considered the 18 metal line indices that can be measured in our spectra, and 
assessed how significant are the corresponding differences $\Delta I$, as follows.
At a fixed age of 14~Gyr and for each  
index, we have estimated the metallicity ${\rm [Fe/H]^{'}}$ 
of the first generation population needed to match the $I$ value of the second generation 
population. If $\Delta I$=0, first and second generation $I$ values are the same, and the 
'correct' metallicity ${\rm [Fe/H]^{'}}$=$-$0.7 of the first generation stars is required to enforce agreement.
A value of $\Delta I$ different from zero requires instead a ${\rm [Fe/H]^{'}}$ 
different from $-$0.7~dex for the first 
generation stars, in order to match the indices of the second generation. 
Table~\ref{tbl1} reports for each metal index the value of $\Delta I$ and the corresponding 
difference $\Delta$[Fe/H] between ${\rm [Fe/H]^{'}}$ and the 'correct' first generation [Fe/H]. 
If we consider as significant $\Delta$[Fe/H] differences above 0.1~dex, 
we find that the only metal indices appreciably affected by the abundance anticorrelations are 
Ca4227, G4300, ${\rm CN_1}$, ${\rm CN_2}$ and NaD. 
The first four indices are sensitive to the C, N, O abundances, while 
NaD is sensitive to the Na and C abundances \citep{trager, korn, prs05, schiavon, gs08}. Not surprisingly, 
${\rm CN_1}$, ${\rm CN_2}$ and NaD are the indices most affected. They display a large  
increase in the second generation population, and to calculate the corresponding $\Delta$[Fe/H] we had to 
perform a linear extrapolation beyond the upper boundary of our available [Fe/H] range.
The values of the Ca4227 and G4300 indices display an opposite behavior, 
decreasing in the second generation population. 
The other metal indices show much less significant variations, when judged on the basis of 
the associated $\Delta$[Fe/H].
All Fe and Mg indices are essentially unaffected, and the same is true for Ca4455, at odds with the behavior 
of Ca4227. The value of ${\rm C_2}$4668 is also very weakly changed, possibly because 
the decrease of both C and O 
in the second generation population affect the index in opposite directions and the resulting net 
variation is very small \citep[see, e.g.][]{trager}. 

\begin{figure*}
\epsscale{1.6}
\plotone{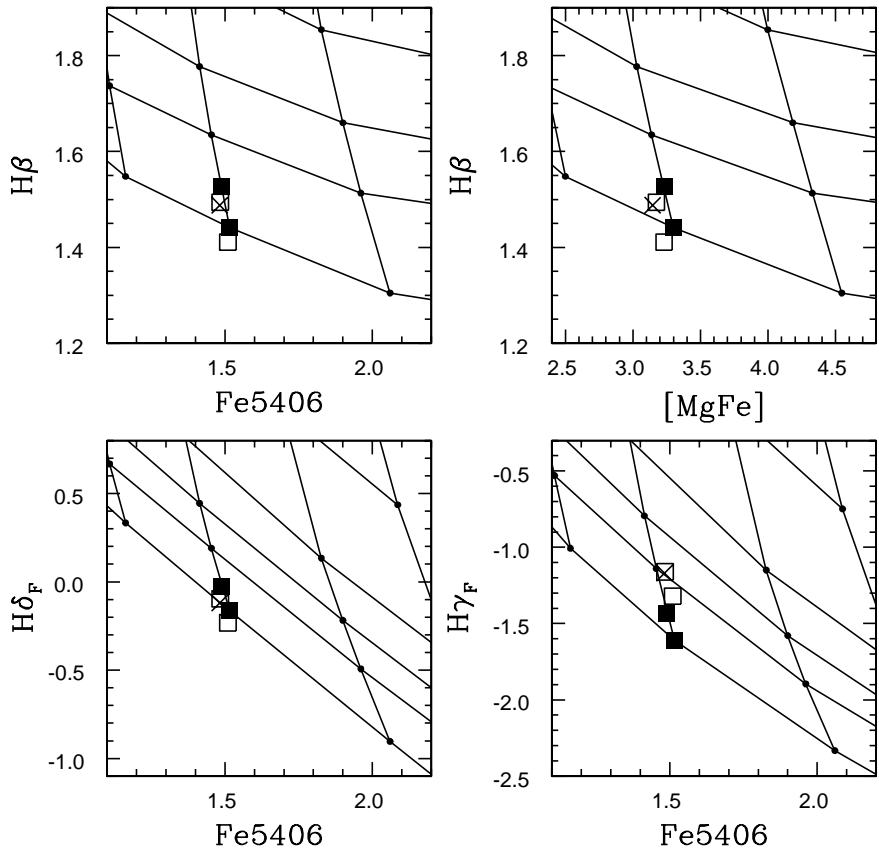}
\caption{Several index-index grids for our reference $\alpha$-enhanced SSPs. 
Metallicity increases from left to right (in the bottom two panels: 
[Fe/H]=$-$1.01, $-$0.70, $-$0.29, +0.05, corresponding to $Z$=0.004, 0.008, 0.01, 0.0198), 
and age increases from top to bottom (in the bottom two panels: t=3, 6, 8, 10, 14~Gyr).
Filled squares denote the 12 and 14 Gyr first generation SSPs, open squares 
the corresponding second generation SSPs with $Y$=0.256, whilst 
crosses denote a 14~Gyr second generation SSP with enhanced He (see text for details).\label{fig1}}
\end{figure*}

We have then investigated the effect of the abundance anticorrelations on the Balmer line indices and 
cluster age estimates, by considering the 
${\rm H\beta}$-Fe5406 and ${\rm H\beta}$-[MgFe] diagrams\footnote{
[MgFe]=$\sqrt{{\rm Mg}b \ \times <{\rm Fe}>}$, with $<{\rm Fe}>$=(Fe5270-Fe5335)/2.} displayed in Fig.~\ref{fig1}. 
These are two powerful diagrams to break the age-metallicity degeneracy and allow the 
estimate of, respectively, age and Fe-abundance, and age and total metallicity $Z$ of SSPs 
\citep[see, e.g.,][and references therein]{w94, thomas05, lw05, tanc05, basti09}.  
Figure~\ref{fig1} confirms the robustness of these diagnostic diagrams. The position of the second generation 
populations on these diagrams is barely distinguishable from their coeval 
first generation counterparts. We have displayed 
in Fig.~\ref{fig1} also the ${\rm H\delta_F}$-Fe5406 and ${\rm H\gamma_F}$-Fe5406 diagrams, that 
essentially confirm the results obtained for the ${\rm H\beta}$ index. 
The ${\rm H\gamma_F}$ index is more affected by the chemical composition 
of the second generation population, and the value of the index increases, 
a behavior opposite to the case of ${\rm H\beta}$ and ${\rm H\delta_F}$.
Ages from ${\rm H\gamma_F}$ are younger by $\sim$2--3~Gyr in second generation stars, but given 
that -- as mentioned at the beginning of this section -- in a real GC   
the variation $\Delta I$ of a generic index is expected to be somewhere within the range 
determined from our model spectra, the `real' 
bias on the ${\rm H\gamma_F}$ ages is probably smaller, at the level of $\approx$1~Gyr.

%%%%%%%%%%%%%%%%%%%%%%%%%%%%%%%%%%%%%%%%%%%%%%%%%%%%%%%%%%%
%                            Figure
%%%%%%%%%%%%%%%%%%%%%%%%%%%%%%%%%%%%%%%%%%%%%%%%%%%%%%%%%%%

%It is interesting to consider also index-index diagrams with one Balmer line age-sensitive 
%index (e.g. ${\rm H\beta}$), and each one of the metal indices that are significantly affected by 
%CNONa anticorrelations, as displayed in Fig.~\ref{fig2}. Due to the non-orthogonal nature of the 
%index-index grids, a large variation of Ca4227, G4300, ${\rm CN_1}$ and ${\rm CN_2}$ indices 
%may cause significant changes in the age estimates, 
%as more clearly visible in case of the  Ca4227 and G4300 indices. 

To summarize, the presence of a second generation of GC stars with unchanged He-content affects appreciably 
only the Ca4227, G4300, ${\rm CN_1}$, ${\rm CN_2}$ and NaD metal indices. 
Very importantly, the variation of Ca4227 goes in the direction to mimic a lower 
Ca abundance, if this index is used as a measure of the Ca content. 
Our results support \citet{lw05} suggestion that 
the effect of a second generation population with CNONa anticorrelations 
may explain the discrepancies they find when comparing 
Ca4227, ${\rm CN_1}$, ${\rm CN_2}$ and NaD index strength (on the Lick IDS system) from their 
$\alpha$-enhanced SSP models, with data for Galactic and M31 GCs.
On the other hand, age, Fe-abundance and $Z$  
inferred from ${\rm H\beta}$-Fe5406 and ${\rm H\beta}$-[MgFe] diagrams (or the ${\rm H\delta_F}$ and,  
to a slightly smaller extent, the ${\rm H\gamma_F}$ counterpart) are confirmed once again to be robust 
and insensitive to the chemical abundance pattern of second generation stars. 

To close our analysis, we consider the case of the 14~Gyr second generation population with enhanced 
He-abundance ($Y$=0.300), also displayed in Fig.~\ref{fig1}. 
The values of all metal indices in this second generation population 
are essentially identical to the $Y$=0.256 case, i.e. they are unaffected 
by the enhancement of He, but the Balmer line indices are changed by the increase of He. The 
values of ${\rm H\beta}$, ${\rm H\delta_F}$ and ${\rm H\gamma_F}$ 
for this population are essentially identical to the 12~Gyr second generation population with $Y$=0.256. 
As a numerical test, we have determined from our new calculations 
all $I$ values for the 14~Gyr second generation population isochrone with 
$Y$=0.256, but using the $Y$=0.300 spectra, and found that both metal and Balmer line indices are identical to 
the values obtained from the spectra with the appropriate $Y$=0.256. 
This means that the changes in ${\rm H\beta}$, ${\rm H\delta_F}$ 
and ${\rm H\gamma_F}$ compared to the $Y$=0.256 case are due to differences in the underlying isochrone  
representative of the He-enhanced second generation population. 
The TO and main sequence of the underlying isochrone are hotter by $\sim$100~K 
compared to the $Y$=0.256 isochrone. Given that the TO mass at fixed age is 
smaller (by about 0.07~${\rm M_{\odot}}$ in our case) in the 
He-enhanced isochrone, and the total mass lost along the RGB is the same (0.11~${\rm M_{\odot}}$) because 
of our assumption of the same mass loss law in all populations, the 
mass evolving along the HB is smaller, hence the typical ${\rm T_{eff}}$ of HB stars is higher, again by $\sim$100~K. 
These higher ${\rm T_{eff}}$ values for TO and HB increase the value of the Balmer line indices.

As a conclusion, the presence of a second generation of GC stars with enhanced He does not modify the 
results obtained for the metal lines 
for the case with $Y$=0.256, but impacts the Balmer line indices, through the ${\rm T_{eff}}$  change of TO and 
(assuming a similar amount of mass lost along the RGB in both first and second generation stars) HB stars. 
At low- and intermediate GC metallicities a small change in the HB evolving mass -- induced for example 
by an enhanced He in second generation stars -- is able produce very blue HB morphologies, 
that can have a huge impact on the cluster ages determined from Balmer line indices 
\citep[see, e.g.,][and references therein]{depr, dfpb00, lyl00, sch04, ps11}.

{\it Acknowledgements: } PC acknowledges the financial support by FAPESP via project 2008/58406-4 
and fellowship 2009/09465-0, and  
is grateful to Fiorella Castelli, Piercarlo Bonifacio and the {\sc kurucz-discuss} mailing list
for the help with ATLAS and SYNTHE codes.

\end{document}